\documentstyle[preprint,aps,eqsecnum]{revtex}

\begin{document}
\def\u{{\cal U}}
\def\v{{\cal V}}
\def\x{{\cal X}}
\def\t{{\cal T}}
\draft
\preprint{gr-qc/9310003  Alberta-Thy-47-93}

\title{Self-dual gravity as a two dimensional theory \\ and
conservation laws}
\author{Viqar Husain}

\address{Theoretical Physics Institute, University of Alberta\\
Edmonton, Alberta T6G 2J1, Canada.}
 \maketitle

\begin{abstract}

Starting from the Ashtekar Hamiltonian variables for general
relativity,
the self-dual Einstein equations (SDE) may be rewritten as evolution
equations
for three divergence free vector fields given on a three dimensional
surface
 with a fixed volume element.

{}From this general form of the SDE, it is shown
 how they may be interpreted as the field equations for a two
dimensional
 field theory. It is further shown that these equations imply an
 infinite number of non-local conserved currents.

 A specific way of writing the vector fields allows an identification
of
 the full SDE with those of the two dimensional chiral
 model, with the gauge group being the group of area preserving
 diffeomorphisms of a two dimensional surface. This gives a natural
 Hamiltonian formulation of the SDE in terms of that of
 the  chiral model. The conservation laws
 using the explicit chiral model form of the equations are also
given.

\end{abstract}
\pacs{4.20.Me, 11.10.Lm, 11.30.Ly}
\vfill
\eject

\section{Introduction}

Integrable systems in two dimensions have been much studied in recent
years and there are  a number of books  that discuss  the
developments
\cite{fadtakh,das,newell}.
Among the main results are the derivations  of  various integrable
non-linear
differential equations from zero curvature conditions.  For example,
it
is known that the KdV   hierarchy of equations may be derived from
an
SL(2,R)  zero curvature condition in 1+1 dimensions using a
particular
parametrization of the  gauge field. It is also known that various
other
integrable equations  may be similarly derived. This is because the
 zero curvature condition is the linearized Lax form
 of a two dimensional non-linear field equation \cite{das}, and it is
 this form which is essential for obtaining conserved quantities and
 proving that they are in involution.

More recent results show that many integrable equations are also
derivable
from self-duality conditions in four dimensions. For example, using
again various parametrizations of an  SL(2,R) gauge field, and this
time
imposing self-duality on the curvatures  followed by a dimensional
reduction
to two dimensions, it has been shown that one can obtain the KdV,
Sine-Gordon,
and non-linear Schrodinger equations \cite{ward,mason,bakas}.
Furthermore,
it has been shown recently that not just the KdV equation but the
entire KdV
hierarchy of  equations  may be obtained from an SL(2,R)
self-duality
condition in 2+2 dimensions \cite{dasgal}.

The two dimensional chiral model based on finite dimensional Lie
groups
 is another example of a two dimensional  integrable system
\cite{fadtakh}.
 One of its field equations is again a zero curvature condition for
the
 relevant group. It is this model, but with an infinite dimensional
Lie group,
 that will be of relevance for the self-dual Einstein equations (SDE)
as
 discussed in this paper. It will turn out, via this analogy that the
SDE
  may also be written as a zero curvature condition and one more
 equation.

  Integrability of the two dimensional models mentioned above is
intimately
connected with the existense of `hidden' symmetries which are
manifested
 through the existence of unexpected conserved quantities, as was
slowly
unravelled first for the KdV equation. Unlike these models, however,
the full Einstein equations contain very few hidden symmetries, apart
from the built in diffeomorphisms.  It has been shown that
the only other symmetries are constant rescalings of the metric
\cite{andtor}.
  This suggests that the vacuum Einstein equations may not be
integrable.
  A further consequence of this for
recent approaches to canonical quantum gravity is the non-existence
of
fully gauge invariant observables  for compact spatial topologies, at
least
for pure gravity, since if such observables existed, they would also
be the
conserved charges associated with hidden symmetries.

However, this state of affairs may not be the case for the much
studied
self-dual Einstein equation
\cite{ludnewtod,pleb}.  There are indications that
this system is entirely integrable \cite{pen} although an infinite
number of
commuting conserved quantities have so far not been constructed.
There
 are a number of interesting results associated with these equations.
 Using a form of the SDE  \cite{ajs} suggested by the Ashtekar
Hamiltonian variables for general relativity \cite{ash}, a
connection
 with the self-dual Yang-Mills equation has been demonstrated
\cite{masnew}:
 the SDE  may be obtained from a 0+1  dimensional  reduction of
 the self-dual Yang-Mills equation by an appropriate choice of gauge
group.
 Another result is that  the field equation for the continuum
 limit  of the Toda model is the same as the SDE for a special
 ansatz for the metric \cite{bakkir}. A further connection with
two dimensional theories has been the derivation of the Plebanski
 equation \cite{pleb} for self-dual metrics from a large N limit of
the
 SU(N) chiral model \cite{park}.

 It is also relatively easy to find solutions of the SDE,
 particularly using their
3+1 form  suggested by the Ashtekar variables.  Using a one-Killing
field
reduction of these equations  (to a  2+1  field theory),  an infinite
class of solutions have been found that are parametrized by elements
of
the $w_\infty$ algebra (which is the algebra of area preserving
diffeomorphisms in two dimensions) \cite{vh}.

These insights, together with the results on
two dimensional models described above raise a number of questions
regarding
the integrability and quantization of the SDE: What is the Lax pair
associated with this equation? Since the Lax pair follows from the
existence
of two distinct  Hamiltonian formulations, what are the two
symplectic forms?
What are the explicit forms of the conserved quantities? Are they in
involution?

 This paper addresses one aspect of these questions, and is concerned
with
elaborating on the connection between the SDE  and
two dimensional theories, with emphasis on conservation laws for the
former.
This will have implications for the existence of fully gauge
invariant
 observables for the self-dual sector of the Einstein equations.

In the next section, we review the construction of non-local
conservation
laws  from the field equations of the chiral model.
Then in section three, we show how the SDE written in the 3+1 form
 of Ashtekar, Jacobson and Smolin may be interpreted as equations for
a two
 dimensional theory. It is shown how this interpretation allows,
using the
 ideas of the second section, an explicit
construction of an infinite number of conserved currents. In section
four
the SDE is written in a form that allows identification with
the chiral model when the gauge group is the group of area
preserving diffeomorphisms of a two dimensional surface. The method
of
section two is again applied to give the conserved currents
explicitly.
The last section contains a discussion and conclusions.

\section{Chiral model and conservation laws}
The chiral field $g(x,t)$ is a mapping from a 2d spacetime  into a
group $g$. The dynamics follows from the Lagrangian density
\begin{equation}
  L = {1\over 2}{\rm Tr}\bigl( \partial_\mu g^{-1}\partial_\nu
g\bigr)
\eta^{\mu\nu}
\end{equation}
where $\eta^{\mu\nu}$ is the flat Minkowski or Euclidean metric.
 The equations of motion are
\begin{equation}
 \partial_\mu (g^{-1}\partial_\mu g)=0
\end{equation}
If we define the Lie algebra valued 1-form  $A_\mu :=
g^{-1}\partial_\mu g$,
 then this equation of motion  becomes
 \begin{equation}
 \partial_\mu A_\mu =0.
 \end{equation}
 Since $A_\mu$ by
definition has a  pure gauge form, it follows that
\begin{equation}
F_{\mu\nu} = \partial_\mu A_\nu - \partial_\nu A_\mu + [A_\mu,A_\nu]
= 0.
\end{equation}
Associated with the gauge field $A_\mu$ there is also the covariant
derivative
 \begin{equation}
 D_\mu = \partial_\mu + [A_\mu,\ ].
\end{equation}
Thus the chiral model describes flat connections $A_\mu$
satisfying  $\partial_\mu A_\mu = 0$. Equations (2.3-2.4) are the
first order
forms of the field equation (2.2).

There exists a simple procedure that demonstrates the existence of
 an infinite number of non-local conserved currents starting
  from equations (2.3-2.4) \cite{bizz}.

 Since the field equation $\partial_\mu A_\mu = 0$ looks like a
conservation law, define the first conserved current to be
\begin{equation}
j_\mu^{(1)}(x) := A_\mu(x).
\end{equation}
 This implies that  there exists
 a function  $f^{(1)}(x)$ such that
 \begin{equation}
 j_\mu^{(1)} = \epsilon_\mu^{\ \nu}\partial_\nu f^{(1)},
 \end{equation}
 where $\epsilon^{\mu\nu}$ is the Levi-Civita tensor.
 We now define the second current
 \begin{equation}
 j_\mu^{(2)}:= D_\mu f^{(1)}.
 \end{equation}
  It is easy to verify that it is conserved:
 \begin{eqnarray}
 \partial_\mu j_\mu^{(2)} &=& D_\mu \partial_\mu f^{(1)}\nonumber \\
 &=& -\epsilon^{\alpha}_{\ \mu}D_\alpha j^{(1)}_\mu =
 -\epsilon^{\alpha}_{\ \mu}D_\alpha D_\mu f^{(0)}\nonumber \\
 &=& -F_{01}f^{(0)}=0,
 \end{eqnarray}
 where the first equality follows because $\partial_\mu A_\mu = 0$,
and
$f^{(0)}=1$ because $j_\mu^{(1)}\equiv A_\mu$.
 Define now the $n$th current by
 \begin{equation}
 j_\mu^{(n)} = D_\mu f^{(n-1)}
 \end{equation}
 Assuming it is conserved, implies there exists a function $f^{(n)}$
such
 that
 \begin{equation}
 j_\mu^{(n)} = \epsilon_\mu^{\ \nu}\partial_\nu f^{(n)}.
 \end{equation}
 It follows that $j_\mu^{(n+1)}$ is conserved:
 \begin{eqnarray}
 \partial_\mu j_\mu^{(n+1)} &=& D_\mu \partial_\mu f^{(n)}\nonumber
\\
 &=& -\epsilon^{\alpha\mu}D_\alpha j^{(n)}_\mu
 =  -\epsilon^{\alpha\mu}D_\alpha D_\mu f^{(n-1)} \nonumber \\
 &=& - F_{01} f^{(n-1)}=0.
 \end{eqnarray}
 Thus the infinite set of currents $j_\mu^{(n)}$ (2.10) are
conserved.
 Since the currents are built up recursively, the associated
conserved
 charges $q^{(n)}$ are non-local. For example
\begin{equation}
 q^{(2)}(t) = \int_{-\infty}^{+\infty} dx\ j^{(2)}_0(x,t)
 = \int_{-\infty}^{+\infty} dx\ (\partial_0 + A_0(x,t))f^{(1)}(x,t),
\end{equation}
 where
\begin{equation}
f^{(1)}(x,t) = \int_{-\infty}^x dx' A_0(t,x')
\end{equation}
follows from integrating (2.7). The integration constant only
 contributes a term proportional to the first conserved charge
$q^{(1)}$.

In the following section it is shown how this procedure may be
applied
 to the self-dual Einstein equation.

\section{Self-dual Einstein equation}
The SDE may be written in a first order form using the Ashtekar
Hamiltonian variables for general relativity \cite{ajs}.

Self-duality is the essential ingredient for this canonical
formulation
and it is natural to ask how the SDE looks in it.
The phase space coordinate is the spatial projection
of the (anti)self-dual part of the spin connection and its conjugate
momentum  is a densitized dreibein. The same is true for
Euclidean or (2,2) signatures, or  complex general relativity, which
are
the cases of interest  for self-dual Riemann curvatures.

 In these Hamiltonian  variables, we would like to know  what is the
 phase space condition  corresponding to the vanishing of the
(anti)self-dual
 part of the four dimensional Riemann curvature. The answer is that
the spatial
 projection of the latter  must be zero.  The vanishing of this
spatial
 projection, when  substituted into Ashtekar's 3+1 evolution
equations leads
 to the new form of the SDE.
 It is straightforward to verify that this condition remains zero
under
 the Hamiltonian  evolution.
 The resulting  equations on four-manifolds $M=\Sigma^3\times R$ may
be
 written in terms of three spatial vector fields $V_i^a$ on
$\Sigma^3$:
\begin{eqnarray}
 Div V^a_i &=& 0  \\
 {\partial V_i^a\over \partial t} &=& {1\over 2} \epsilon_{ijk}[
V_j,V_k]^a,
 \end{eqnarray}
where the divergence is defined with respect to a constant  auxiliary
density
 and the right hand side of (3.2) is the Lie bracket.
The self-dual
four metrics  are constructed from  solutions of these equations
using
\begin{equation}
g^{ab} = ({\rm det} V)^{-1}  [ V^a_i V^b_j \delta^{ij}  + V_0^a V_0^b
].
\end{equation}
Here  $i,j,k...=1,2,3$ label the vector field, $a,b,...$ are abstract
vector
indices,
$V_0^a$ is the vector field that is used to perform the 3+1
decomposition,
and
$ \partial V_i^a/\partial t \equiv V_0^b\partial_b V_i^a$.
 The time derivative in (3.2) can be written in the more general form
 $[V_0,V_i]^a$.
(For details of the derivation of these equations the reader is
referred  to \cite{ajs} where they were originally derived, or the
review in \cite{vh}).

Our starting point will be the SDE in the form (3.1-3.2).
We first rewrite equation (3.2) in a form similar to that suggested
by
Yang \cite{yang} for the self-dual Yang-Mills equation
 \begin{equation}
 F_{ab}={1\over 2}\epsilon_{ab}^{\ \ cd}F_{cd}
 \end{equation}
  on a complex manifold. Replacing the (local) complex flat
coordinates
 $x_0,...,x_3$ by the linear combinations $t=x_0+ix_1$,
 $u=x_0-ix_1$, $x=x_2-ix_3$ and  $v=x_2+ix_3$,
 equation (3.4) becomes
 \begin{eqnarray}
  F_{tx}& = &F_{uv}=0 \\
 F_{tu}& + &F_{xv}=0.
 \end{eqnarray}

For the SDE, defining in a similar way
\begin{eqnarray}
\t &=& V_0+iV_1 \ \ \ \ \ \ \ \ \ \u= V_0-iV_1 \nonumber \\
\x &=& V_2-iV_3 \ \ \ \ \ \ \ \ \ \v=V_2+iV_3,
\end{eqnarray}
the evolution equations (3.2) become
\begin{equation}
[\t,\x] = [\u,\v]=0
\end{equation}
\begin{equation}
 [\t,\u ] + [\x,\v]=0,
\end{equation}
where the vector indices have been suppressed. This shows a rather
direct
analogy between  the self-dual Yang-Mills and Einstein equations,
namely,
the Yang-Mills curvatures are replaced by the Lie brackets of the
vector
fields. A similar analogy has been noted in a related way in
ref. \cite{masnew}.

We will now show that equations (3.8-3.9) give rise to an infinite
number of
 conservation laws in a manner similar to that for the chiral model
 described in the last section. We fix (locally) a coordinate system
$(t,x,p,q)$ and using the gauge freedom, fix in these coordinates
\begin{equation}
\t^a = ({\partial\over \partial t})^a \ \ \ \ \ \ \ \ \ \
\x^a = ({\partial\over \partial x})^a,
\end{equation}
with $\u,\v$
 arbitrary except that they be divergence free with respect to the
 volume form defined by the local coordinates:
 $\omega = dt\wedge dx \wedge dp \wedge dq$.
 Equations (3.8-3.9) then become
\begin{equation}
 {\partial \u \over \partial t} +
 {\partial \v \over \partial x} = 0,
 \end{equation}
 \begin{equation}
 [\u,\v] = 0.
  \end{equation}
These are to be compared with equations (2.3-2.4) of the chiral
model.
 The former has the form of a continuity
equation on the
two dimensional $(x,t)$ plane where there is a flat 2-d
background metric $\delta_{\mu\nu}$, $\mu,\nu, ...=x,t$. Define
a two dimensional vector field valued 1-form whose components
are the vector fields $\u,\v$:
\begin{equation}
 A_\mu = (\u dt  + \v dx)_\mu.
 \end{equation}
Equivalently, $A_\mu$ is a dyad whose components are vector fields.
In this notation, we can rewrite equation (3.9) as
$\delta^{\mu\nu}\partial_\mu A_\nu=0$
and so the first conserved current, by analogy
with equation (2.6), is $J_\mu^{(1)}:= A_\mu$.
Thus there exists a vector field $\eta^{(1)}$ such that
\begin{equation}
J_\mu^{(1)}=\epsilon_\mu^{\ \nu}\partial_\nu \eta^{(1)} .
\end{equation}

To avoid  notational confusion, we emphasize that for the analogy
with
the chiral model equations, we are working in a fixed (local)
coordinate
system $t,x,p,q$, part of which $(t,x)$ is being interpreted as a two
dimensional `background'  spacetime, and that the vector fields
$\u,\v$ have
components in this background, as well as in the `internal' $(p,q)$
space.
In addition, the vector fields $\u,\v$ are themselves the $(x,t)$
components
of the vector field valued gauge field $A_\mu$ as defined in equation
(3.13).
Thus for example, (3.14) is an equation relating the {\it components}
of the
vector fields $J_\mu^{(1)}$ and $\eta^{(1)}$ in the given coordinate
system.

Again following the analogy with the chiral model, we now define a
second current and show that it is conserved. Since the analogy with
the chiral model isn't exact, this definition is different from
(2.8).
Let
\begin{equation}
  J_\mu^{(2)} := [A_\mu, \eta^{(1)}]
\end{equation}
 We have
 \begin{eqnarray}
  \delta^{\mu\nu}\partial_\mu J_\nu^{(2)} &=&
 \delta^{\mu\nu}\partial_\mu [A_\nu,\eta^{(1)}]
  = \delta^{\mu\nu} [A_\nu,\partial_\mu\eta^{(1)}] \nonumber \\
 &=&
 -\delta^{\mu\nu} [A_\nu,J^{(1)}_\alpha]\epsilon^\alpha_{\ \mu}
 = -2[\u,\v]=0
\end{eqnarray}
by equation (3.12). Assuming now a conserved current
 \begin{equation}
  J_\mu^{(n)}:= [A_\mu,\eta^{(n-1)}] =
 \epsilon_\mu^{\ \nu}\partial_\nu \eta^{(n)}
 \end{equation}
it is easy to  show that
\begin{equation}
 J_\mu^{(n+1)}:= [A_\mu, \eta^{(n)}]
 \end{equation}
is conserved:
 \begin{eqnarray}
 \delta^{\mu\nu}\partial_\mu J_\nu^{(n+1)} &=&
 \delta^{\mu\nu} [A_\nu,\partial_\mu\eta^{(n)}] \nonumber \\
  &=& -\epsilon^{\mu\nu} [A_\nu, J_\mu^{(n)}]=
 -\epsilon^{\mu\nu} [A_\nu, [A_\mu,\eta^{(n-1)}]] \nonumber \\
  &=& -[\u,[\v,\eta^{(n-1)}]] + [\v,[\u,\eta^{(n-1)}]] \nonumber \\
 &=& -[[\u,\v],\eta^{(n-1)}]=0
 \end{eqnarray}

Thus we have shown that equations (3.11-3.12) imply the conservation
of the infinite number of vector field valued currents
$ J_\mu^{(n)}$ defined in equation (3.17).  All these currents are
clearly
independent, as
 for the chiral model case,  by the recursive construction of
equation
 (2.10) or (3.18).
 (We remind the reader that the two components $\mu=0,1$ of
 $A_\mu$ or $J_\mu^{(n)}$ corresponding to
 the $t,x$ coordinates, are themselves four dimensional vector
 fields, and all the above equations hold for each component of the
 vector fields in the coordinate system.)

The corresponding conserved charges are four dimensional vector
fields.
The components of the first two charges are  given in  terms
 of the integrals of the components of the vector field $\u$ using
 \begin{eqnarray}
 Q^{(1)}(t) &=& \int dxdpdq\ J_0^{(1)}=\int dxdpdq\ \u  \\
 Q^{(2)}(t) &=& \int dxdpdq\ J_0^{(2)}=\int dxdpdq
 \ [\u,\int^{x}dx'dpdq\ \u(t,x',p,q)]
 \end{eqnarray}
This shows the non-local nature of the charges.
These are however only formal expressions, since for their evaluation
for particular metrics, the integration limits will depend  on the
`spatial' topologies.

\section{Self-duality equations as a chiral model}

In the last section we introduced a coordinate system and showed
that the SDE implies an infinte set of conserved currents
 by using an analogy with the two dimensional chiral model. The
 SDE (3.11-3.12) in this coordinate system do not however take
 exactly the chiral model form.

In this section we use a specific but general form for the
vector fields $\u,\v$ and show that the SDE  may be written
{\it exactly} as the chiral model equations (2.3-2.4), with the gauge
group
being the
group of area preserving diffeomorphisms of a two dimensional
surface.

Before doing this we first give, for later comparison, the connection
 between the equations
 (3.11-3.12) and the Plebanski equation \cite{pleb,cmn}. The vector
fields
 $\x,\t$ as chosen in the above coordinate system are already
divergence
 free with respect to the volume form $\omega$. We now take the
 vector fields $\u,\v$ to have the following form, such that
 they are divergence free, and are given in terms of one function
 $\Omega(t,x,p,q)$:
 \begin{eqnarray}
 \u &=& -\Omega_{xq}{\partial \over \partial p} +
 \Omega_{xp}{\partial \over \partial q} \\
 \v &=& \Omega_{tq}{\partial \over \partial p} -
  \Omega_{tp}{\partial \over \partial q}.
  \end{eqnarray}
 where the subscripts denote partial differentiation. This
 identically satisfies equation (3.11), and equation (3.12) gives
\begin{equation}
\Omega_{xp}\Omega_{tq} - \Omega_{xq}\Omega_{tp} = h(x,t) \nonumber,
\end{equation}
 where $h(x,t)$ is an arbitrary function. With the change of
variables
  $q\rightarrow qh(x,t)$, this becomes Plebanki's first equation
 \begin{equation}
\Omega_{xp}\Omega_{tq} - \Omega_{xq}\Omega_{tp} = 1.
 \end{equation}
 Furthermore, equation (3.3) for the metric leads to the line element
\begin{equation}
ds^2 = \Omega_{tp} dtdp + \Omega_{tq}dtdq + \Omega_{xp}dxdp
 + \Omega_{xq}dxdq
 \end{equation}

 We now show how the SDE  may be written as a chiral model
 field equation. Again working in the above coordinate system, and
 with the same choice (3.10) for $\t,\x$, we take the following
 divergence free form\footnote{This form is similar to, but more
 general than that used by the author previously in ref. \cite{vh},
 where one-Killing field reductions of the  self-duality equations
 are discussed.}
  for $\u,\v$ in terms of the two functions $A_0(t,x,p,q)$ and
 $A_1(t,x,p,q)$:
 \begin{eqnarray}
 \u^a &=& ({\partial \over \partial t})^a + \alpha^{ba}\partial_b A_0
\\
 \v^a &=& ({\partial \over \partial x})^a + \alpha^{ba}\partial_b
A_1,
 \end{eqnarray}
 where $\alpha^{ab} = (\partial/\partial p)^{[a}\otimes
 (\partial/\partial q)^{b]}$  is the antisymmetric tensor that is the
 inverse of the two form $(dp\wedge dq)_{ab}$ in the $p,q$ plane.
 Substituting (4.6-4.7) into (3.11-3.12) gives
 \begin{eqnarray}
 \alpha^{ab}\partial_b \bigl[\partial_0 A_1 - \partial_1 A_0 +
 \{A_0,A_1\}\bigr] &=& 0\\
 \alpha^{ab}\partial_b \big[\partial_0A_0 + \partial_1 A_1\bigr] = 0.
 \end{eqnarray}
 where the bracket on the left hand side of equation (4.8) is the
 Poisson bracket with respect to $\alpha^{ab}$:
 \begin{equation}
 \{A_0,A_1\} := \alpha^{ab}\partial_aA_0\partial_bA_1=
 \partial_p A_0 \partial_q A_1
 -\partial_q A_0 \partial_p A_1,
 \end{equation}
 and $\partial_0,\partial_1$ denote partial derivatives with respect
to
 $t,x$ etc. Equations (4.8-4.9) imply that the terms in their
 square brackets are equal to two arbitrary functions of $t,x$, which
 we write as
 \begin{eqnarray}
  \partial_0 A_1 - \partial_1 A_0 +
 \{A_0,A_1\} &=& \partial_0 F(t,x) +\partial_1 G(t,x)\\
 \partial_0A_0 + \partial_1 A_1 &=& \partial_1 F(t,x) - \partial_0
G(t,x),
\end{eqnarray}
(where $F,G$ are arbitrary.)
With the redefinitions
\begin{equation}
a_0(t,x,p,q):= A_0 + G \ \ \ \ \ \ a_1(t,x,p,q) := A_1 - F,
\end{equation}
(4.11-4.12) become
\begin{eqnarray}
\partial_0 a_1 - \partial_1 a_0 +
 \{a_0,a_1\} &=& 0 \\
 \partial_0 a_0 + \partial_1 a_1 &=& 0.
 \end{eqnarray}
 These are precisely the chiral model equations (2.3-2.4) on the
 $x,t$ `spacetime', with $p,q$ treated as coordinates on an internal
 space. The commutator has been replaced by the Poisson bracket
(4.10).
 The gauge group
 is therefore the group of diffeomorphisms that preserve
$\alpha^{ab}$ on
 the internal space. (Note that the redefinitions (4.13) do not alter
the
 vector fields $\u,\v$ in equations (4.6-4.7))

 Unlike in the last section,  the procedure of section 2 for
constructing
conservation laws applies more directly to this case.  The only
difference  is that the Lie group is now infinite dimensional.
Denoting the pair $a_0,a_1$ by $a_\mu$, we have the covariant
derivative
\begin{equation}
 D_\mu = \partial_\mu + \{a_\mu,\ \}.
\end{equation}
The first current is as before $J_\mu^{(1)}:= a_\mu$, which implies
that there exists a function $g(t,x,p,q)$ such that
$J_\mu^{(1)}=\epsilon_\mu ^{\ \nu}\partial_\nu g^{(1)}$. The
$n$th conserved current is defined by
\begin{equation}
J_\mu^{(n)} := D_\mu g^{(n-1)}
\end{equation}
where like equation (2.7), $g^{(n-1)}$ is defined by
$J_\mu^{(n-1)}=\epsilon_\mu^{\ \nu}\partial_\nu g^{(n-1)}$. The
inductive
 proof in section 2 of the conservation of these currents goes
through
 unaltered.

The conserved charges are now integrals of the functions $a_0$. The
first two are
\begin{eqnarray}
Q^{(1)}(t) &=& \int dxdpdq\ a_0 \\
Q^{(2)}(t) &=& \int dxdpdq\ \bigr[\partial_0 g^{(1)} +
\{a_0,g^{(1)}\}\bigl]
\end{eqnarray}
where $g^{(1)} = \int^{x} dx'dpdq\ a_0(t,x',p,q)$.

As contrasted with the last section where no fixed form of the vector
 fields $\u,\v$ is used, here we have the form (4.6-4.7) which leads
to a
specific  internal gauge group. This group can fixed by choosing the
`internal' space to be some specific two-dimensional manifold
$\Sigma$
and then one can try to solve the two dimensional equations
(4.14-4.15).
Since $a_\mu$ is valued in the Lie algebra of the area preserving
diffeomorphisms, for compact internal
spaces,  such as $T^2$, the integrals over $p,q$ can be done
explicitly
by working in a specific basis for the Lie algebra. Such bases have
been studied for a number of two surfaces \cite{poprom}. We may
further
choose the space coordinatized by $x$ to be $S^1$ and seek
 solutions periodic in $x$. This will then allow
an explicit evaluation of all the conserved charges for compact
spatial three manifolds of topology $S^1\times\Sigma$.

There is a further question regarding the relation between the
full SDE (3.11-3.12) and the chiral model equations (4.14-4.15)
derived from them. How general is the form (4.6-4.7) for the vector
fields?  We can see that the
 two first order equations for $a_0,a_1$
 are equivalent to a single second
 order equation for a function $\Lambda(t,x,p,q)$ obtained by writing
 \begin{equation}
 a_0=\partial_1\Lambda  \ \ \ \ \ \ \
 a_1=-\partial_0\Lambda
 \end{equation}
 This solves equation (4.15) while equation (4.14) becomes
\begin{equation}
\Lambda_{tt} + \Lambda_{xx} +
\Lambda_{xp} \Lambda_{tq} - \Lambda_{xq} \Lambda_{tp}=0,
\end{equation}
(where the subscripts denote partial  derivatives.)
This is one equation for a function of all the spacetime coordinates
and therefore doesn't represent any reduction in the local degrees of
freedom for self-dual metrics. Using equation (3.3) the line element
is
\begin{eqnarray}
ds^2 &=&
  - dt ( \Lambda_{tp}dp + \Lambda_{tq}dq )
 - dx ( \Lambda_{xp}dp + \Lambda_{xq}dq )\nonumber \\
 & & + {1\over \{\Lambda_t,\Lambda_x\}}\bigl(
     ( \Lambda_{xp}dp + \Lambda_{xq}dq )^2
    + ( \Lambda_{tp}dp + \Lambda_{tq}dq )^2\bigr)
\end{eqnarray}

Equation (4.21) is essentially another way of writing the SDE, which
is
different from Plebanski's two equations.\footnote{Another
alternative
to the Plebanski equation has also been discussed recently
\cite{grant}.
This approach also starts from the Ashtekar-Jacobson-Smolin equations
but the remaining development is different from that given here, as
is
the final equation.}
  That it has the same content is seen most
simply by comparing the forms of the vector fields $\u,\v$ (4.1-4.2),
which give the first Plebanski equation, with the following final
form of
the vector fields (obtained by substituting (4.20) into (4.6-4.7)),
which give the alternative equation (4.21):
\begin{eqnarray}
\u &=&  {\partial\over \partial t} -
 \Omega_{xq}{\partial \over \partial p} +
 \Omega_{xp}{\partial \over \partial q} \\
\v &=& {\partial\over \partial x}+
\Omega_{tq}{\partial \over \partial p} -
 \Omega_{tp}{\partial \over \partial q}.
\end{eqnarray}
The forms of the vector fields (4.1-4.2) and (4.23-4.24) clearly have
 the same functional content.

The  usefulness of equation (4.21) is its direct connection with the
chiral model, which leads easily to the conservation laws and the
Hamiltonian formulation. The latter is given in ref. \cite{fadtakh}
for finite dimensional groups, and this immediately generalizes
to the infinite dimensional case.

\section{Conclusions and Discussion}

There are two main results presented in this paper. The first
provides
a simple way to view the SDE  as the field equation of a two
dimensional
theory, the chiral model with gauge group the group of area
preserving diffeomorphisms of a two dimensional surface. The second
result
is an explicit construction of an infinite number of conserved
currents for
the SDE, which was presented here in two different ways
in sections 3 and 4. \footnote{A recent
paper also discusses non-local conservation laws for the
SDE  using a completely different twistor approach
\cite{strach}. The method makes use of the connection between the
null surfaces corresponding to self-dual spaces and the
Plebanski equation. It would be of interest to compare this to
 the approach given above.}

The first result has been noted before in a different
context in \cite{park} where the inverse scattering form of the
 chiral model equations were related to Plebanski's first equation.
 Here, on the other hand,  we have seen that the explicit chiral
 model equations (4.14-4.15)
 may be derived from the Ashtekar-Jacobson-Smolin form of the
 SDE, and that its second order form leads to equation (4.21),
 which is different from the two Plebanski heavenly equations.

The second result shows that, unlike the full Einstein equations, the
self-dual sector is rather rich in conservation laws. This has
implications for the quantum theory for this sector, since in the
canonical approach to quantum gravity, one of the goals is to
identify
fully gauge invariant phase space functionals (`observables'), which
are
to be represented as Hermitian operators on the Hilbert space. This
has
been a stumbling block for full Einstein gravity where no
such observables are known. The self-dual sector may be
viewed as perhaps the largest possible midi-superspace (albeit with
the  wrong signature metric), but the results given  here show that
its
phase space has an infinite number of observables.

A further implication of these results is that, unlike the Plebanski
heavenly equations, equation (4.21) has the
advantage that it has an obvious Hamiltonian formulation - that of
the
two dimensional chiral model. In particular it should be possible
to rewrite the conserved charges in terms of Hamiltonian variables
and compute explicitly their Poisson brackets. An important question
for
 integrability is whether these charges are in involution.
But whether or not they are in involution, we still have an infinite
number of observables at hand for canonical quantization.

All known integrable two dimensional models  have two Hamiltonian
structures. The two symplectic forms  provide an elegant way to
generate
all the conserved charges. An interesting question in this context
 is whether the SDE has two symplectic forms, and
whether the conserved charges given here may be derived using these
forms
in the standard way \cite{das}. The answer to this may be provided by
asking what the two symplectic forms are for the chiral model.

 An important result from the study of two dimensional models
is that the transfer matrix, which is the path ordered exponential
 of the connection in the zero curvature form of the model, provides
 a way of constructing the conserved quantities. The connection in
the zero curvature condition normally depends on the spectral
parameter,
and therefore the trace of the transfer matrix depends on it as well.
This trace serves as the generating function for the conservation
laws.
Namely, the coefficients of powers of the spectral parameter in the
expansion of the trace are the conserved charges \cite{fadtakh,das}.
(This procedure has recently been applied to the two Killing field
reductions of the Einstein equations \cite{nmbs}.)

On the other hand, here we have derived the conserved charges, which
have
a resemblance to terms in the expansion of such a trace, (the
 expansion of the Wilson loop for example.) However, it is not known
how
 to construct the holonomy associated with an infinite dimensional
group, let
alone take its trace. In this regard it may be possible to invert
the transfer matrix method  of obtaining conserved charges, and use
the
latter obtained here to get an expansion for the holonomy associated
with
an infinite dimensional Lie group. Work on this and on the canonical
 quantization of the SDE  using the above ideas is in progress
 \cite{nmvh}.

\acknowledgements
This work was supported by NSERC of Canada.

 \end{document}